# Higgs boson self coupling from two-loop analysis

*H. A. Alhendi*[1,2*], *T. Barakat*[1**], *and I. Gh. Loqman*[1†]

*1 Department of Physics and Astronomy*

*College of Science, King Saud University*

*P.O. Box 2455, Riyadh 11451, Saudi Arabia*

*2 National Center for Mathematics and Physics. KACST.*

*P.O. Box 6086, Riyadh 11442, Saudi Arabia*

## Abstract

The scale invariant of the effective potential of the standard model at two-loop is used as a boundary condition under the assumption that the two-loop effective potential approximates the full effective potential. This condition leads with the help of the renormalization group functions of the model at two-loop to an algebraic equation of the scalar self coupling with coefficients that depend on the gauge and the top quark couplings. It admits only two real positive solutions. One of them, in the absence of the gauge and top quark couplings, corresponds to the non-perturbative ultraviolet fixed point of the scalar renormalization group function and the other corresponds to the perturbative infrared fixed point. The dependence of the scalar coupling on the top quark and the strong couplings at two-loop radiative corrections is analyzed.

---

*Electronic address: alhendi@ksu.edu.sa / hhalhendi@gmail.com

** Electronic address: tbarakat@ksu.edu.sa

† Electronic address: ibrahim_lok@yahoo.com

## I. Introduction

The standard model requires, through Higgs mechanism [1], the existence of neutral scalar Higgs boson that survives after spontaneous breakdown of the electroweak symmetry. In this mechanism the weak gauge bosons and the charged fermions as well as the scalar Higgs itself acquire their masses from the asymmetry of the vacuum [2]. At tree level the scalar Higgs boson self-coupling ($\lambda$), and thus the Higgs boson mass ($m_H$), is a free parameter of the model and its determination self-consistently poses a challenging problem. Many efforts, theoretically and experimentally, have been devoted to it.

One of the primary goals of LHC is to search for the standard model Higgs boson. Recent high energy experiments excluded the region $m_H \leq 114.4$ GeV [3] and electro-weak precession data excluded the region 160 GeV < $m_H$ < 170 GeV [4]. The region 114.4GeV < $m_H$ < 160GeV is a favored one, while the region $m_H$ > 180 GeV, though disfavored, is not ruled out. The current fits lead to130 GeV < $m_H$ < 260 GeV [5]. Recently the global fit obtained by the Gfitter Group showed that, the 95% CL allowed range for the complete fit (including the direct searches) is [114,153] GeV, and above this range only the values between 180 GeV and 224 GeV are not yet excluded at 3 standard deviations or more [6].

Several papers have presented upper bounds based on avoiding triviality of pure $\varphi^4$-theory [7], lower bounds on the bases of vacuum stability [8], and constraints based on some theoretical assumptions. The requirement, for example, that the coefficient of the quadratic divergent term of the square of the scalar Higgs boson mass at one-loop is equal zero leads to the well-known Veltman condition [9], the assumption that the one-loop effective potential effectively represents the full effective potential leads to an algebraic equation that fixes, in the absence of fermion contributions, the scalar self-coupling [10], and from the assumption that the ratio of the scalar coupling to the square of the top quark coupling is scale invariant allows to fix the Higgs boson mass at one and two-loop[11].

In the standard model for the potential ($m^2 < 0$, $\lambda > 0$)

$$V^{(0)} = \frac{1}{2}m^2\varphi^2 + \frac{\lambda}{4!}\varphi^4 \,, \qquad (1)$$

the two-loop renormalization group function $\beta_\lambda$ for pure $\varphi^4$-theory is given by [12]:

$$\beta_\lambda = \beta_\lambda^{(1)} + \beta_\lambda^{(2)} = \frac{1}{16\pi^2}(4\lambda^2) + \frac{1}{(16\pi^2)^2}(-\frac{26}{3}\lambda^3) \ . \quad (2)$$

This equation admits two constant solutions: the perturbative infrared fixed point $\lambda$=0, and the nonperturbative ultraviolet fixed point $\lambda_{\mathrm{UV}}$=72.9. It is expected, similar to the one loop case [13], that the perturbative fixed point is removed in the presence of interactions with gauges and fermions while the nonperturbative one persists [14,15]. We show that this is the case, under the assumption that the two-loop effective potential approximates the full effective potential. This assumption leads to an algebraic equation that fixes the scalar self coupling in the perturbative region. The ultraviolet fixed point, where perturbation expansion breaks down [14, 15], is shown to occur in the region where all other couplings, but the scalar coupling, are neglected in comparison.

The full effective potential $V(\varphi(\mathrm{t}))$, which is constructed from the one-particle-irreducible Green's functions at zero external momenta [16], satisfies, in the mass-independent scheme of Weinberg-t'Hooft [17], the renormalization group equation $\mathrm{d}V(\varphi(\mathrm{t}))/\mathrm{dt}=\mathrm{D}V(\varphi)=0$, where D is a first order partial differential operator that depends on the renormalization group functions of the model and t is an arbitrary running parameter related to the renormalization scale μ (t=ln(μ/μ(0)). This equation is just a statement that $V$ is scale invariant $V(\varphi(\mathrm{t}))=V(\varphi(0))$ [16]. When perturbative expansion to a given order is employed to V one may take this scale invariance as a boundary condition for the solution of the first order partial differential equation. In the present work we use this boundary condition at two-loop to obtain a relation that fixes the scalar coupling in terms of the gauge and top quark couplings. This relation expresses the validity of the scale invariant of the two-loop effective potential.

This paper is organized as follows; in the following section we calculate the effective potential at two loop approximation using the renormalization group equation in the mass independent subtraction scheme. In section. III we impose the scale invariance on the two-loop effective potential that allows to obtain a condition that expresses the validity of the scale invariant at two-loop, which depends on the renormalization group functions at two-loop. This condition leads to an algebraic equation of degree five in $\lambda$, which has only one perturbative real positive solution, and one nonperturbative real positive solution. In section. IV the perturbative solution is used to estimate the mass of the Higgs boson at one and two loop approximation, the dependence of the Higgs boson mass on the strong interaction coupling $\alpha_{\mathrm{S}}$ and the top quark mass $m_t$ is discussed, in this section we apply the matching condition to estimate the physical mass of the Higgs boson. Finally in section. V we give our conclusion.

## II. Calculation of the effective potential at two-loop

In the mass-independent subtraction scheme of Weinberg-t'Hooft, the scale invariance of the full effective potential of a renormalizable quantum field theory leads, under infinitesimal transformation, to the renormalization group equation (RGE) of the effective potential [17]:

$$\left(\mu\frac{\partial}{\partial\mu}+\delta_p\beta_p\frac{\partial}{\partial\lambda_p}+\gamma\varphi_c\frac{\partial}{\partial\varphi_c}\right)V_{eff}(\varphi_c)=0, \qquad (3)$$

where $\mu$ is a renormalization mass parameter and $\gamma$ is the anomalous dimension, $\lambda_p$=($m^2$,λ,$g$, $g'$, $g_3$, $g_t$) , $g$ ,$g'$ , $g_3$ are the SU(2) , the U(1), and the SU(3) gauge couplings respectively, $g_t$ is the top quark coupling and $\delta_p$ equals $m^2$ for mass coupling and 1 otherwise, and $\beta_p$ is the renormalization group function:

$$\beta_p=\frac{d\lambda_p}{dt}, \quad \text{with} \quad t=\ln\frac{\mu}{\mu(0)}.$$

In the loop expansion, the effective potential is given by [14]:

$$V_{eff}=V^{(0)}+\sum_{n=1}\hbar^{(n)}V^{(n)}, \qquad (4)$$

following the procedure of ref. [18], we can write the following recursion equation for the effective potential up to order *n*:

$$\mu\frac{\partial}{\partial\mu}V^{(n)}+D_nV^{(0)}+D_{n-1}V^{(1)}+\cdots+D_1V^{(n-1)}=0, \qquad (5)$$

where the differential operator $D_n$ is given by:

$$D_n=\delta_p\beta_p^{(n)}\frac{\partial}{\partial\lambda_p}+\gamma^{(n)}\varphi_c\frac{\partial}{\partial\varphi_c}, \qquad n=1,2,3,\ldots \qquad (6)$$

and $V^{(n)}$ is the n-loop contribution to $V$, and $\beta_p^{(n)}$ is the n-loop contribution to the renormalization group function:

$$\beta_p=\sum_{n=1}\hbar^{(n)}\beta_p^{(n)}. \qquad (7)$$

To find the one-loop effective potential we use eqs. (5), (6), (7) and the zero-loop (the tree level) potential of eq. (1), which give:

$$V^{(1)}=\frac{1}{48}a_1\varphi^4\ln\frac{\varphi^2}{M^2}, \quad where \quad a_1=\beta_\lambda^{(1)}+4\gamma^{(1)}\lambda, \qquad (8)$$

and for the 2-loop we substitute eqs. (1) and (8) into eq. (6) to find:

$$V^{(2)} = \frac{1}{48}\ a\ \varphi^4 \ln\frac{\varphi^2}{M^2} + \frac{1}{192}\ b\ \varphi^4 \left(\ln\frac{\varphi^2}{M^2}\right)^2 ,$$

$$where\ \ a = a_1\gamma^{(1)} + a_2\ ,\ with\ \ a_2 = \beta_\lambda^{(2)} + 4\gamma^{(2)}\lambda\ , \tag{9}$$

$$\text{and}\ \ b = \beta_\lambda^{(1)}\frac{\partial\beta_\lambda^{(1)}}{\partial\lambda} + 8\ \beta_\lambda^{(1)}\gamma^{(1)} + 16\ \lambda\ (\gamma^{(1)})^2 + G, \tag{10}$$

with

$$G = \beta_{g'}^{(1)}\frac{\partial\beta_\lambda^{(1)}}{\partial g'} + 4\lambda\beta_{g'}^{(1)}\frac{\partial\gamma^{(1)}}{\partial g'} + \beta_g^{(1)}\frac{\partial\beta_\lambda^{(1)}}{\partial g} + 4\lambda\beta_g^{(1)}\frac{\partial\gamma^{(1)}}{\partial g} + \beta_{g_t}^{(1)}\frac{\partial\beta_\lambda^{(1)}}{\partial g_t} + 4\lambda\beta_{g_t}^{(1)}\frac{\partial\gamma^{(1)}}{\partial g_t}\ . \tag{11}$$

Therefore, the full effective potential approximated to two-loop effective potential becomes:

$$V = V^{(0)} + V^{(1)} + V^{(2)} = A_0 + A_1\varphi^4\ln\frac{\varphi^2}{\mu^2} + A_2\varphi^4(\ln\frac{\varphi^2}{\mu^2})^2\ , \tag{12}$$

$$\text{where}\ \ A_0 = \frac{1}{2}m^2\varphi^2 + \frac{\lambda}{4!}\varphi^4\ ,$$

$$A_1 = \frac{1}{48}(a_1(1+\gamma^{(1)}) + a_2)\ , \tag{13}$$

$$\text{and}\quad A_2 = \frac{1}{192}\left(\frac{8}{k}\lambda\ \beta_\lambda^{(1)} + 4a_1\gamma^{(1)} + G\right). \tag{14}$$

The renormalization group functions used in this work are [12]:

$$\gamma^{(1)} = \frac{1}{4k}\left(3g'^2 + 9g^2 - 12g_t^2\right)\ ,$$

$$\beta_\lambda^{(1)} = \frac{1}{k}\left(4\lambda^2 - \lambda\left(3g'^2 + 9g^2 - 12g_t^2\right) + \frac{9}{4}g'^4 + \frac{9}{2}g'^2g^2 + \frac{27}{4}g^4 - 36g_t^4\right),$$

$$\beta_{g_t}^{(1)} = \frac{1}{k}(\frac{9}{2}g_t^3 - 8g_3^2g_t - \frac{9}{4}g^2g_t - \frac{17}{12}g'^2g_t)\ ,$$

$$\beta_g^{(1)} = \frac{-19}{6k}g^3\ ,\ \beta_{g'}^{(1)} = \frac{41}{6k}g'^3\ ,\ \beta_{g_3}^{(1)} = \frac{-7}{k}{g_3}^3\ ,$$

$$\gamma^{(2)} = -\frac{1}{6k^2}\lambda^2 - \frac{1}{k^2}(\frac{431}{96}g'^4 + \frac{9}{16}g'^2g^2 - \frac{271}{32}g^4 + \frac{85}{24}g'^2g_t^2 + \frac{45}{8}g^2g_t^2 + 20g_3^2g_t^2 - \frac{27}{4}g_t^4)\ ,$$

$$\beta_\lambda^{(2)} = \frac{1}{k^2}(-\frac{26}{3}\lambda^3 + \lambda^2\left(6g'^2 + 18g^2 - 24{g_t}^2\right) + \lambda\left(\frac{629}{24}g'^4 + \frac{39}{4}g'^2g^2 + \frac{85}{6}g'^2{g_t}^2 - \frac{73}{8}g^4 + \frac{45}{2}g^2{g_t}^2 + 80{g_3}^2{g_t}^2 - 3{g_t}^4\right) + 180{g_t}^6 - 192g_3^2g_t^4 - 16g'^2g_t^4 - \frac{27}{2}g^4g_t^2 + 63g'^2g^2g_t^2 - \frac{57}{2}g'^4g_t^2 - \frac{379}{8}g'^6 - \frac{559}{8}g'^4g^2 - \frac{289}{8}g'^2g^4 + \frac{915}{8}g^6)\ ,$$

$\beta_{g_t}^{(2)} = \frac{1}{k^2}\, g_t(-12g_t^4 + g_t^2\left(\frac{131}{16}g'^2 + \frac{225}{16}g^2 + 36g_3^2 - 2\lambda\right) + \frac{1187}{216}g'^4 - \frac{3}{4}g^2g'^2 + \frac{19}{9}g'^2g_3^2 - \frac{23}{4}g^4 + 9g^2g_3^2 - 108g_3^4 + \frac{1}{6}\lambda^2)$ ,

$\beta_{g}^{(2)} = \frac{1}{k^2}g^3(\frac{3}{2}g'^2 + \frac{35}{6}g^2 + 12g_3^2 - \frac{3}{2}g_t^2)$ ,

$\beta_{g'}^{(2)} = \frac{1}{k^2}g'^3(\frac{199}{18}g'^2 + \frac{9}{2}g^2 + \frac{44}{3}g_3^2 - \frac{17}{6}g_t^2)$ ,

$\beta_{g_3}^{(2)} = \frac{1}{k^2}g_3^3(\frac{11}{6}g'^2 + \frac{9}{2}g^2 - 26g_3^2 - 2g_t^2))$ ,

and $k = 16\pi^2$. (15)

## III. A condition from the scale invariance at two-loop

As stated in the introduction, when a perturbative expansion of the effective potential to a given order is employed, the boundary condition

$V(\varphi(t),\lambda_p(t),\mu(t))=V(\varphi, \lambda_p,\mu)$ , (16)

where $\varphi(t) = z\varphi$, $\mu(t)=e^t\,\mu$ leads to a restriction on the coupling constants that expresses the validity of this scale invariant to this order.

Employing these transformations in the 2-loop effective potential (12), we obtain the following conditions:

$A_1(t)z^4 + 4A_2(t)z^4\ln\left(\frac{z}{e^t}\right) = A_1(0)$ , (17)

and

$A_2(t)z^4 = A_2(0)$ . (18)

Differentiating eq. (13), using eq. (14) and the definition of $\gamma$ :

$\gamma = \frac{1}{z}\frac{dz}{dt}$ , (19)

we obtain the following relation:

$\frac{d\,A_1}{dt} + 4A_1\gamma + 4A_2(\gamma - 1) = 0$ , at t=0. (20)

Substituting $A_1$ and $A_2$ from eqs. (13), (14) and using eqs. (9), (10), (11) we finally obtain:

$$\left(-1+\gamma^{(1)}+\gamma^{(2)}\right)\left(\frac{1}{2\pi^2}\lambda\beta_\lambda^{(1)}+4\gamma^{(1)}\left(\beta_\lambda^{(1)}+4\gamma^{(1)}\lambda\right)+G\right)+$$
$$4\left(\gamma^{(1)}+\gamma^{(2)}\right)\left(\left(1+\gamma^{(1)}\right)\left(\beta_\lambda^{(1)}+4\gamma^{(1)}\lambda\right)+\left(\beta_\lambda^{(2)}+4\gamma^{(2)}\lambda\right)\right)+$$
$$\frac{d}{dt}\left(\left(1+\gamma^{(1)}\right)\left(\beta_\lambda^{(1)}+4\gamma^{(1)}\lambda\right)+\left(\beta_\lambda^{(2)}+4\gamma^{(2)}\lambda\right)\right)=0\,, \qquad (21)$$

where G is as given in eq.(11).

Substituting eq. (11) and the one and two-loop renormalization group functions from eq. (15) into eq. (21), and after simplifying, we obtain an equation of degree five in $\lambda$:

$$\lambda^5+C_1\lambda^4+C_2\lambda^3+C_3\lambda^2+C_4\lambda+C_5=0\,, \qquad (22)$$

where $C_n$ (n=1,2,..,5) are functions in the couplings $g'$, $g$, $g_3$, $g_t$, and the explicit expressions of them are given in the appendix.

## IV. The Higgs boson mass

In the absence of the couplings $g$, $g'$, $g_3$, and $g_t$ , eq.(22) has only two solutions, one of them corresponds to the ultraviolet fixed point solution $\lambda_{UV}$=72.9, in which the perturbative expansion breaks down, and the other is the infrared fixed point solution $\lambda$=0. This result can be taken as a check of the validity of the scale invariant of the effective potential at two-loop corrections. In the presence of these couplings the non-perturbative solution continues and we are left with only one real positive perturbative solution which we use it to analyze the dependence of the scalar coupling for a range of values of $g_3$ and $g_t$.

Our input parameters are obtained from the relations:

$$g=\frac{2M_W}{v}\,,\ g'=\sqrt{\frac{4M_Z^2}{v^2}-g^2}\,,\ \ g_t=\sqrt{2}\,\frac{m_t}{v}\,,\text{ with } v=246.2\text{ GeV}$$

$M_Z$ = 91.19 GeV, $M_W$ = 80 GeV,

$g'(M_Z)$=0.35554 , $g(M_Z)$=0.64988.

We take the mass of the top quark in the range 160 GeV≤ $m_t$ ≤176 GeV, and the strong coupling $\alpha_S$ in the range $0.1073 \leq \alpha_S \leq 0.1202$ [19] .

In table. I we present our results for the scalar Higgs coupling in the above ranges of $m_t$, and $\alpha_S$ .

The Higgs boson mass, at zero external momentum, is defined by [16]:

$$m_H^2=\frac{d^2V}{d\varphi^2}\Big|_{\varphi=v}\,, \qquad (23)$$

where $v$ is the vacuum expectation value which has the expansion up to two-loop:

$$v = v_0 + \hbar v^{(1)} + \hbar^2 v^{(2)}, \tag{24}$$

and satisfies

$$\frac{dV}{d\varphi}\Big|_{\varphi=v} = 0\ . \tag{25}$$

The Higgs boson mass at tree level is:

$$m_H{}^2 = \frac{1}{3}\lambda\, v_0{}^2, \tag{26}$$

where $v_0$=246.2 GeV,

at one-loop, using eq. (6), is:

$$m_H{}^2{}_{1-loop} = [\frac{1}{3}\lambda + \frac{1}{4}a_1 + \frac{1}{3}a_1 ln\frac{v_1}{\mu}]v_1{}^2, \tag{27}$$

where $v_1 = v_0 + \hbar v^{(1)}$,

and at two-loop, using eq.(9), is:

$$m_H{}^2{}_{2-loop} = [\frac{1}{3}\lambda + 12A_1 + 8A_2 + 8(A_1 + 3A_2)(ln\frac{v_2{}^2}{\mu^2}) + 8A_2(ln\frac{v_2{}^2}{\mu^2})^2]v_2{}^2\ , \tag{28}$$

where $v_2 = v_1 + \hbar^2 v^{(2)}$.

A comparison between the tree level potential and two loop effective potential for $\alpha_S$=0.1161, $m_t$=171 GeV is shown in Fig.1, while for $\alpha_S$=0.118, $m_t$=171 GeV is shown in Fig. 2. Here it should be noted that all other figures with different values of $\alpha_S$ and $m_t$ show similar trends. As can be inferred from these figures the two-loop contribution to the effective potential almost cancels the one-loop contribution in the region in between the positions of the two minima of the potential and as a result the position of the vacuum is slightly shifted (by less than one percent)[20]. For each of the obtained value of the scalar coupling the positivity of the effective potential at two-loop is checked from the condition $V(\varphi) > 0$, for $|\varphi| > \varphi^*$, where $\varphi^*$ is the non trivial positive solution of $V(\varphi^*)=0$.

When the perturbative real positive solutions of eq.(22) is substituted in the tree-level eq.(26), one-loop eq.(27) and the two- loop eq.(28) we find the mass of the Higgs boson at $M_Z$ to be :

224.56 GeV $\leq m_{H\text{-tree}} \leq$ 236.85 GeV,

219.62 GeV $\leq m_{H\text{-1loop}} \leq$ 236.03 GeV,

222.92 GeV $\leq m_{H\text{-2loop}} \leq$ 238.95 GeV.

The dependence of $m_{H\text{-1loop}}$ and $m_{H\text{-2loop}}$ on the top quark mass and the strong coupling $\alpha_S$ is shown in the Fig. 3, which is in a qualitative agreement with ref .[14] ( see Fig. 3a ) and ref.[21]( see Fig. 3) in the region considered.

Using the matching conditions, the initial value $m_H(M_Z)$ is related to the physical Higgs mass $M_H$ to one-loop order at the scale $M_Z$ [22, 19, 21]

by :

$$m_H^2 = M_H^2(1 + \delta(M)) \qquad (29)$$

where $$\delta(M) = \frac{1}{16\,\pi^2}\frac{M_Z^2}{v^2}[\xi f_1(\xi, M) + f_0(\xi, M) + \xi^{-1} f_{-1}(\xi, M)] \qquad (30)$$

where $\xi \equiv \frac{M_H^2}{M_Z^2}$ and $f_1(\xi,M)$, $f_0(\xi,M)$, $f_{-1}(\xi,M)$ are given in the appendix.

Inserting the input parameters, we find that the physical mass of the Higgs boson: 221.9 GeV $\leq M_H \leq$ 240.1 GeV. This result is in the expected range of the recent global fit to all precision electroweak data [5,6]. Now we compare our result with some results of other authors employing different approaches. In ref.[23] a value of about 218 GeV for the Higgs boson mass ($m_H$)is obtained from using renormalization-group methods to include all leading-logarithm contributions to the effective potential, and in ref.[24] a value of $m_H$ = 221 GeV from consideration of scalar-field theory projection of the effective potential. In ref.[25] using the renormalization group improved effective potential and results from ref.[26] they obtained an upper bound of $m_H$ around 196 GeV. In ref.[27] from their estimate of the top quark mass using Goldberger- Teller relation and Veltman condition they found $m_H$ around 317 GeV. The assumption made in ref. [28] that the scalar coupling equals the top quark coupling led to a value of $m_H$ around 348 GeV. Finally, an estimated value of $m_H$=700 GeV has been obtained in ref. [29] from the assumption that the ratio of the Higgs boson mass to the vacuum expectation value is a cutoff-independent.

## V. Conclusion

In this paper we have used the scale invariant of the effective potential at two-loop as a boundary condition .This condition which expresses the validity of the scale invariant of effective potential at two-loop, results in an algebraic equation of degree five. In the physical ranges of the top quark mass and the strong gauge coupling it admits only two real positive solutions. In the absence of the gauge and the top quark couplings one of these two solutions corresponds to the ultraviolet fixed point of the scalar renormalization group function, where perturbation theory breaks down, and the other one lies in the perturbative regime. Our findings support, in general, several other works. Of these , the two-loop contribution almost cancels the contribution of the one-loop contribution in the region between the minima of the potential and thus the vacuum expectation value of the scalar Higgs boson is slightly shifted by renormalization at one- loop and two-loop quantum corrections. The other finding is that the variation of the Higgs boson mass as a function of the top quark mass initially decreases in the region considered. Finally we have made a comparison of our result with some results of other authors employing different approaches.

*TABLE. I. Values of the scalar coupling $\lambda$ obtained from eq(22), for different values of the top quark mass $m_t$ (GeV) and the strong coupling $\alpha_S$ , column 4 gives the Higgs boson mass at tree level (eq(26)),column 5 at 1-loop (eq(27)), column 6 at 2-loop (eq(28)),and column 7 the physical Higgs boson mass from eq(29).*

| $\alpha_S$, $g_3$ | $m_t$(GeV) | $\lambda$ | $m_H$(tree) | $m_H$-1Lp | $m_H$-2Lp | $M_H$-1L |
|---|---|---|---|---|---|---|
| 0.1073, 1.1612 | 160 | 2.45170 | 224.56 | 223.93 | 226.56 | 227.16 |
| | 162 | 2.45764 | 224.83 | 223.60 | 226.30 | 226.71 |
| | 164 | 2.46305 | 225.08 | 223.21 | 225.98 | 226.20 |
| | 166 | 2.46793 | 225.30 | 222.77 | 225.62 | 225.64 |
| | 168 | 2.47224 | 225.50 | 222.27 | 225.20 | 225.02 |
| | 170 | 2.47597 | 225.67 | 221.70 | 224.72 | 224.33 |
| | 172 | 2.47910 | 225.81 | 221.08 | 224.18 | 223.59 |
| | 174 | 2.48159 | 225.92 | 220.38 | 223.58 | 222.77 |
| | 176 | 2.48343 | 226.00 | 219.62 | 222.92 | 221.90 |
| 0.1127, 1.19005 | 160 | 2.54599 | 228.83 | 229.14 | 231.88 | 232.71 |
| | 162 | 2.55308 | 229.15 | 228.88 | 231.70 | 232.32 |
| | 164 | 2.55966 | 229.45 | 228.56 | 231.46 | 231.87 |
| | 166 | 2.56571 | 229.72 | 228.20 | 231.18 | 231.38 |
| | 168 | 2.57121 | 229.96 | 227.77 | 230.84 | 230.82 |
| | 170 | 2.57615 | 230.19 | 227.29 | 230.45 | 230.21 |
| | 172 | 2.58052 | 230.38 | 226.75 | 230.00 | 229.54 |
| | 174 | 2.58428 | 230.55 | 226.15 | 229.49 | 228.81 |
| | 176 | 2.58741 | 230.69 | 225.48 | 228.92 | 228.01 |
| 0.1161, 1.20787 | 160 | 2.60418 | 231.43 | 232.31 | 235.13 | 236.10 |
| | 162 | 2.61198 | 231.78 | 232.09 | 234.99 | 235.75 |
| | 164 | 2.61927 | 232.10 | 231.82 | 234.81 | 235.34 |
| | 166 | 2.62604 | 232.40 | 231.50 | 234.57 | 234.88 |
| | 168 | 2.63227 | 232.68 | 231.13 | 234.28 | 234.38 |
| | 170 | 2.63795 | 232.93 | 230.69 | 233.93 | 233.80 |
| | 172 | 2.64306 | 233.16 | 230.20 | 233.53 | 233.17 |
| | 174 | 2.64758 | 233.35 | 229.65 | 233.08 | 232.49 |
| | 176 | 2.65149 | 233.53 | 229.03 | 232.57 | 231.74 |
| 0.118, 1.21772 | 160 | 2.63634 | 232.86 | 234.05 | 236.92 | 237.97 |
| | 162 | 2.64453 | 233.22 | 233.85 | 236.80 | 237.63 |
| | 164 | 2.65222 | 233.56 | 233.61 | 236.64 | 237.25 |
| | 166 | 2.65938 | 233.87 | 233.31 | 236.43 | 236.81 |
| | 168 | 2.66601 | 234.17 | 232.96 | 236.16 | 236.32 |
| | 170 | 2.67209 | 234.43 | 232.56 | 235.85 | 235.78 |
| | 172 | 2.67761 | 234.67 | 232.09 | 235.48 | 235.17 |
| | 174 | 2.68254 | 234.89 | 231.57 | 235.05 | 234.51 |
| | 176 | 2.68688 | 235.08 | 230.98 | 234.56 | 233.79 |
| 0.1202, 1.22901 | 160 | 2.67320 | 234.48 | 236.03 | 238.95 | 240.10 |
| | 162 | 2.68184 | 234.86 | 235.86 | 238.87 | 239.79 |
| | 164 | 2.68997 | 235.22 | 235.65 | 238.73 | 239.43 |
| | 166 | 2.69759 | 235.55 | 235.38 | 238.55 | 239.02 |
| | 168 | 2.70467 | 235.86 | 235.06 | 238.31 | 238.56 |
| | 170 | 2.71122 | 236.14 | 234.68 | 238.03 | 238.03 |
| | 172 | 2.71720 | 236.40 | 234.24 | 237.69 | 237.45 |
| | 174 | 2.72261 | 236.64 | 233.75 | 237.29 | 236.82 |
| | 176 | 2.72743 | 236.85 | 233.20 | 236.84 | 236.13 |

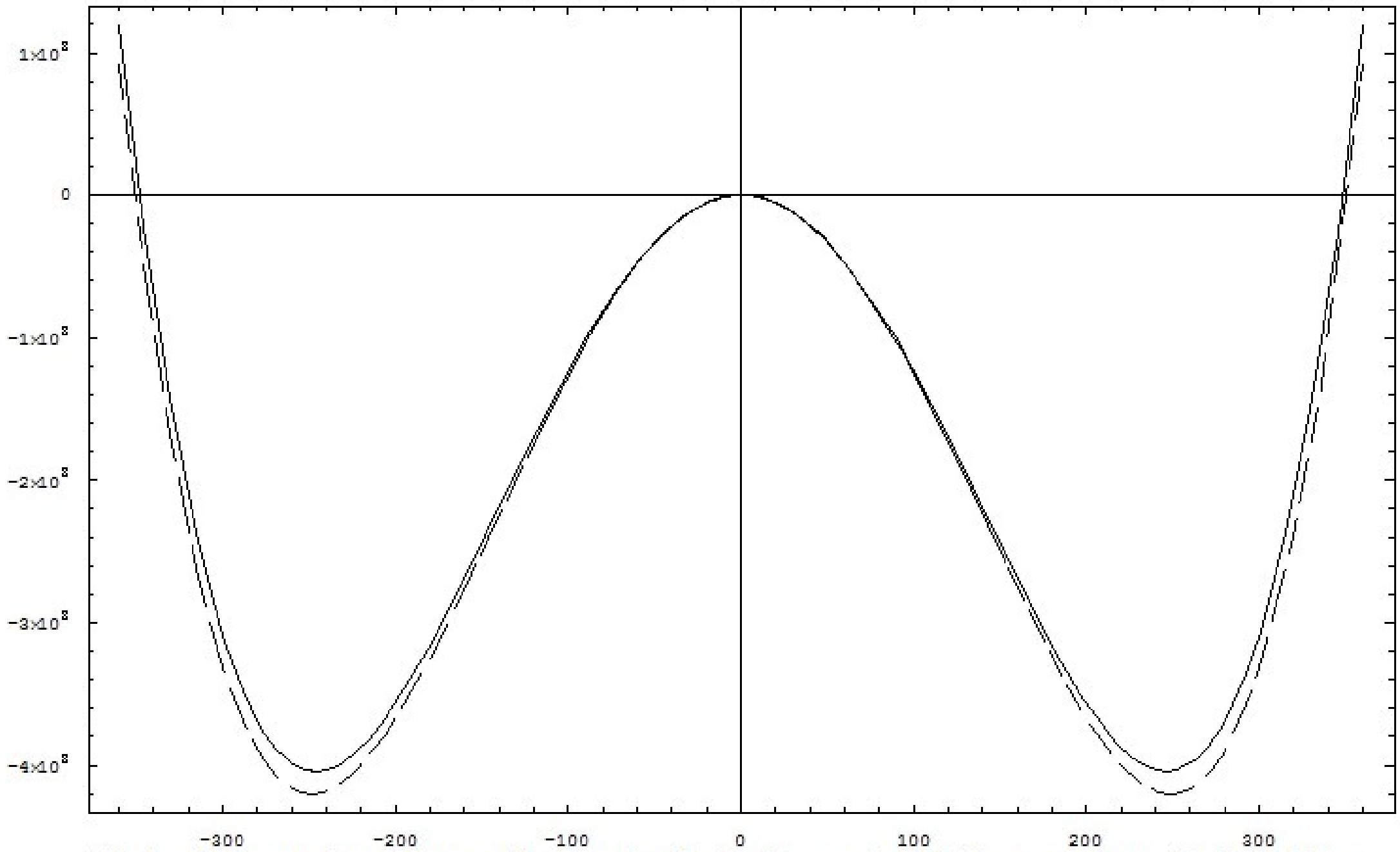

Fig.1. A comparison between the tree level effective potential (the contineous line) and the 2-loop effective potential (the dashed line) for $a_S$=0.1161 and $m_t$ =171 GeV

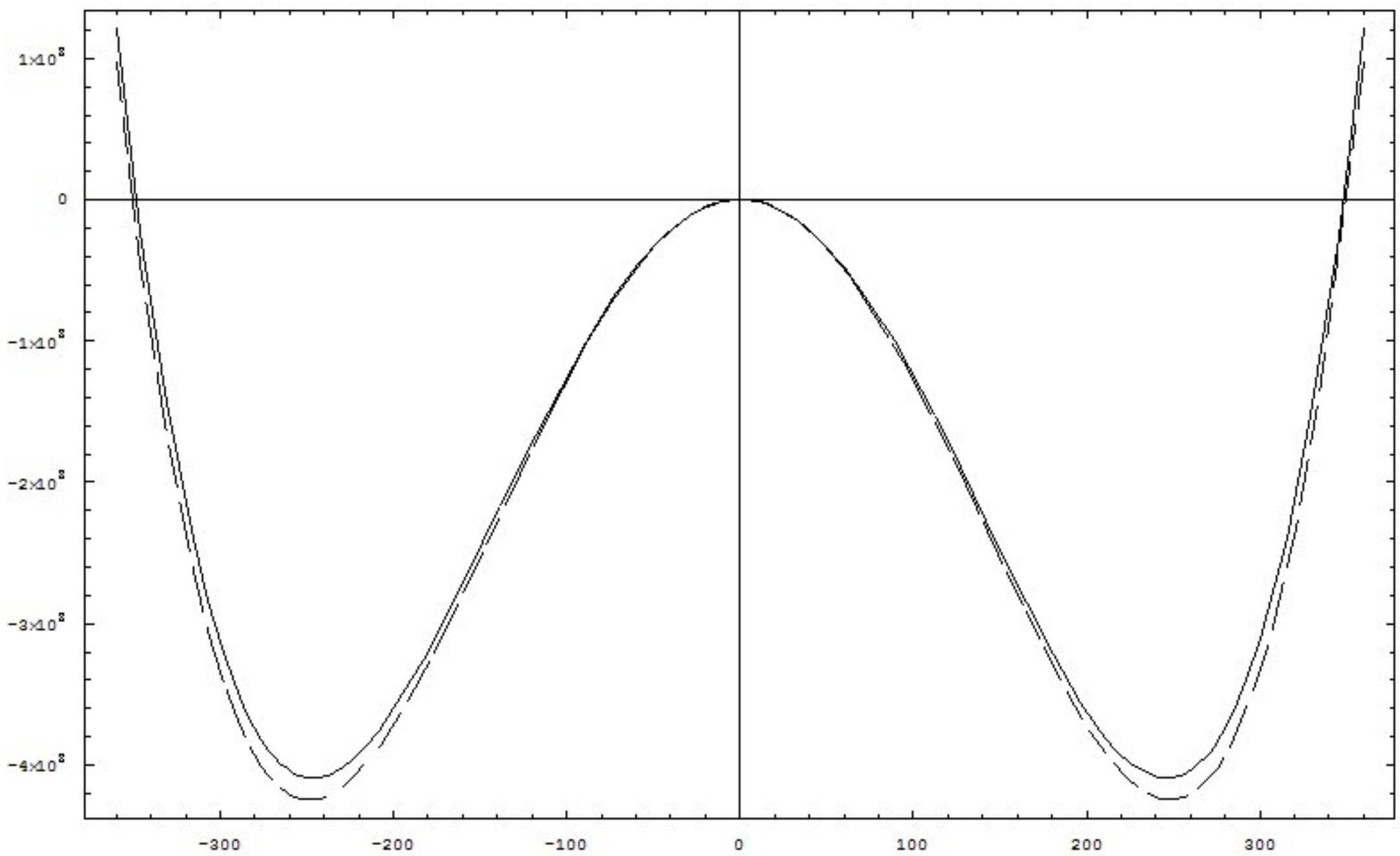

Fig.2. A comparison between the tree level effective potential (the contineous line) and the 2-loop effective potential (the dashed line) for $a_S$=0.1180 and $m_t$ =171 GeV

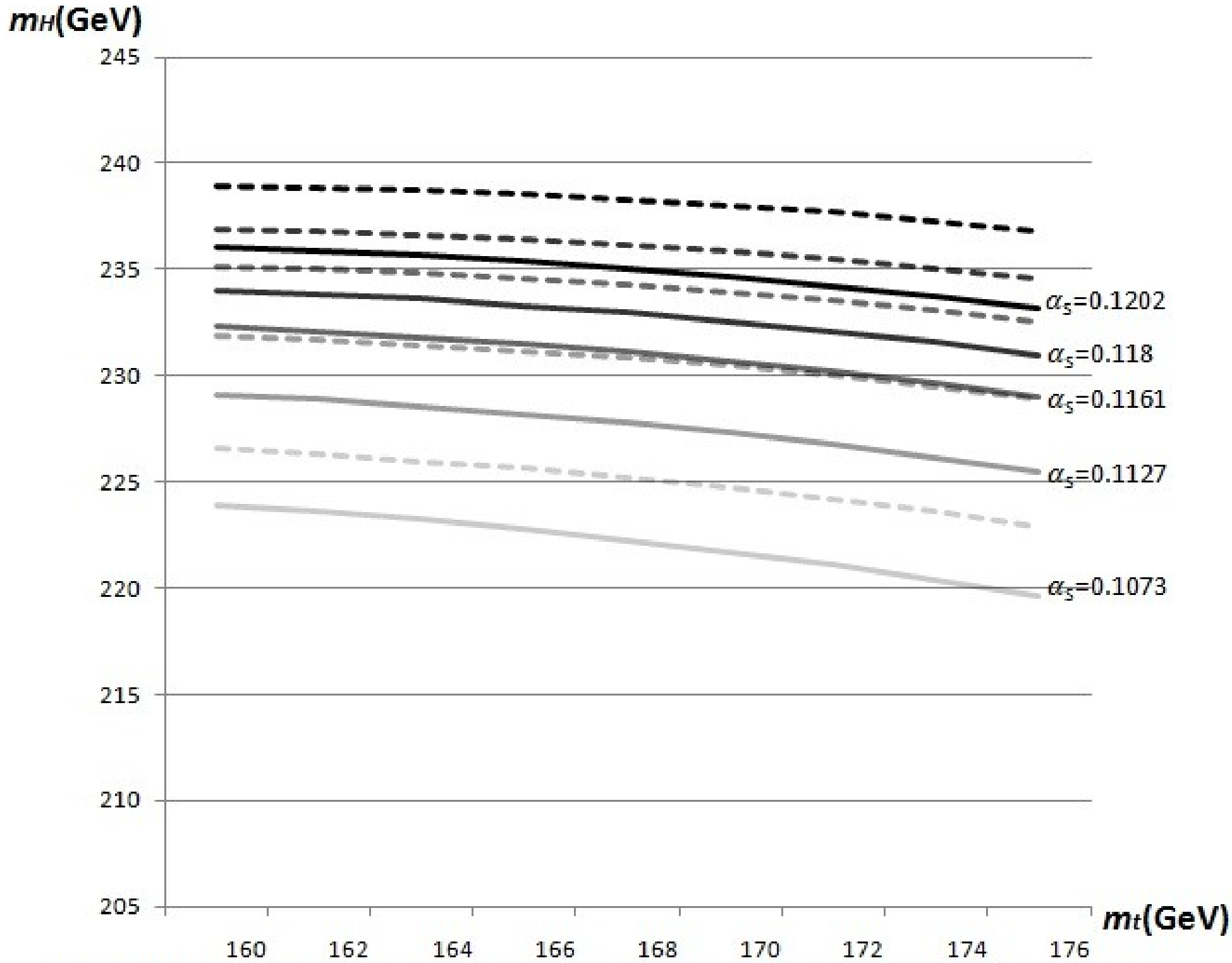

Fig .3. The Higgs boson mass $m_H$ (GeV) verses the top quark mass $m_t$ (GeV) for different values of the strong coupling $\alpha_S$ . The solid line corresponds to the 1Loop case and the dashed line corresponds to the 2Loop ,the darkness of the lines increases with increasing $\alpha_S$

*Refferences*

[1] P. W. Higgs, Phys. Rev. Lett. **13** (1964) 508; ibid. Phys. Rev. **145** (1966) 1156.

[2] S. Weinberg, Phys. Rev. Lett. **19** (1967) 1264; A. Salam in " Elementary Particle Theory" , ed. N. Svartholm, Almqvist and Wiksells, Stockholm (1968) 367.

[3] The ALEPH, DELPHI, L3 and OPAL Collaboration, and LEP Working Group for Higgs Boson Search, Phys. Lett. **B 565** (2003) 61.

[4] Tevatron New Phenomena and Higgs Working Group for the CDF and D0 Collaborations, FERMILAB-PUB-09-060-E, "Combined CDF and DZero Upper Limits on Standard Model Higgs-Boson Production with up to 4.2 fb-1 of Data ": hep-ex/0903.40010

[5] Abdelhak Djoudi, Phys. Rep. **457** (2008) 1

[6] H. Flächer *et al.*, Eur. Phys. J. **C 60**, (2009) 543 (hep-ph/0811.0009); updated resultstaken from http://cern.ch/gfitter; J. Ellis, J. R. Espinosa, G. F. Giudice, A. Hoecker and A. Riotto, Phys. Lett. **B679** (2009) 369 .

[7] L. Maiani, G. Parisi, and R. Petronzio, Nucl. Phys. **B 136** (1978) 115; R. Dashen and H. Neuberger, Phys. Rev. Lett. **50** (1983) 1897; M. A. Beg, C. Panagiotakopolus and A. Sirlin, Phys. Rev.Lett. **52** (1984) 883

[8] N. Cabibo, L. Maiani, G. Parisi and R. Petronzio, Nucl. Phys. **B158** (1972) 295; M. Lindner, Z. Phys. **C31** (1986) 295; M. Sher, Phys. Rept. **179** (1989) 273; M. Lindner, M. Sher and H. W. Zaglauer, Phys. Lett. **B 228** (1989) 139; M. Sher, Phys. Lett**. B317** (1993) 159; Addendum-ibid, **B331** (1994) 448; G. Altarelli and G Isidori, Phys. Lett. **B 337** (1994) 141; J. A. Casas, J. R. Espinosa, M. Quiros, Phys. Lett. **B382** (1996) 374.

[9] M. Veltman, Acta Phys. Pol. **B12** (1981) 437; I. Jack and D. R. T. Jones, Phys. Lett. **B345** (1990); Nucl. Phys**. B342** (1990) 127; M. Capdequi Peyranere, J. C. Montero and G. Moultaka, Phys. Lett. **B260** (1991) 138; P. Osland, T. T. Wu, Z. Phys. **C55** (1992) 585; E. Ma, Phys. Rev. **D47** (1993) 2143; G. Lopez Castro and J. Pestiean, Mod. Phys. Lett. **A10** (1995) 1155; M.Chaichian, R. Gonzalez Felipe and K Huitu, " On Quadratic Divergences and the Higgs Mass", Phys.Lett. B363 (1995) 101-105: hep-ph/9509223.

[10] H. Al-Hendi, M. Ozer and M. O. Taha, Z. Phys. **C36** (1987) 629.

[11] J. Pasupathy, Mod. Phys. Lett **A15** (2000) 1605; B. Ananthanarayan and J. Pasupathy, " Higgs Mass in the Standard Model from Coupling Constant Reduction", Int.J.Mod.Phys. **A17** (2002) 335: hep-ph/0104286.

[12] M. Machacek and M.T. Vaughn, Nucl. Phys**. B249** (1985) 70; C. Ford, I. Jack and D. R. T. Jones, Nucl.Phys. **B387** (1992) 373[Erratum-ibid. **B504** (1997) 551; " The Standard Model Effictive Potencial at Two Loops" : hep-ph/011190; M. X. Luo and Y. Xiao, Phys. Lett. **90** (2003) 011601.

[13] D. J. E. Callaway, Nucl. Phys. B233 (1984) 189.

[14] B. Grzadkowski and M. Linder, Phys. Lett. **B178** (1986) 81.

[15] J. S. Lee and J. K. Kim, Phys. Rev**. D53** (1996) 6689.

[16] S. Coleman and E. Weinberg, Phys. Rev. **D7** (1973) 1888.

[17] G. t' Hooft, Nucl. Phys. **B61** (1973) 455; S. Weinberg, Phys. Rev. **D8** (1973) 3497.

[18] H. Alhendi, Phys. Rev **D37** (1988)3749 [ Erratum Phys. Rev. **D40** (1989) 683].

[19] Particle Data Group Collaboration, C. Amsler *et al*., " Review of particle physics," Phys. Lett. **B 667** (2008) 1.

[20] H. Arason, D. J. Castano, B. Keszthelyi, S. Mikaelian, E. J. Piard, P. Ramond, and B. D. Wright, Phys. Rev. **D46** (1992) 3945.

[21] T. Hambye and K. Riesselmann, Phys. Rev. **D55** (1997) 7255.

[22] A. Sirlin and Zucchini, Nucl. Phys. **B266** (1986) 389.

[23] V. Elias , R. B. Mann, D, G, C. McKeon , and T. G. Steele, Phys. Rev. Lett. **91** (2003) 251601.

[24] F. A. Chishtie, V. Elias, and T.G. Steele, Int.J.Mod.Phys.**A20** (2005) 6241, " Startling equivalences in the Higgs-Goldstone sector between radiative and lowest-order conventional electroweak symmetry breaking": hep-ph/0502044.

[25] P. P. Pal and S. Chakrabarty, Acta. Phys. Pol.**B40** (2009) 1645.

[26] P. Kielanowski and S. R. Juarez W, Phys. Rev.**D72** (2005) 096003.

[27] M. D. Scadron, R. Delbourgo, and G. Rupp, J.Phys.**G32** (2006) 735, " Constituent Quark Masses and the Electroweak Standard Model": hep-ph/0603196.

[28] V.A. Bendnyakov, N. D. Giokaris, and A. V.Bednyakov, Phys.Part.Nucl.**39** (2008) 13, "On Higgs mass generation mechanism in the standard model" : hep-ph/0703280.

[29] P. Cea and L. Cosmai,"The Higgs boson: from the lattice to LHC": hep-ph/0911.5220.

## Appendix

$C_1 = \frac{1}{420864}(-476928g'^2 - 1430784g^2 + 1893888g_t^2 - 3170304\pi^2)$

$C_2 = \frac{1}{420864}(-1240512g'^4 + 148608g'^2g^2 + 1068480g^4 - 1648128g'^2g_t^2 - 4022784g^2g_t^2 - 3686400g_3^2g_t^2 + 1963008g_t^4 + 4202496g'^2\pi^2 + 12607488g^2\pi^2 - 16809984g_t^2\pi^2)$

$C_3 = \frac{1}{420864}(2997216g'^6 + 5102496g'^4g^2 + 3044448g'^2g^4 - 4057344g^6 + 304128g'^4g_3^2 + 746496g^4g_3^2 - 727968g'^4g_t^2 + 116640g^4g_t^2 + 654336g'^2g_3^2g_t^2 + 1741824g^2g_3^2g_t^2 + 8957952g_3^4g_t^2 - 22464g'^2g_t^4 - 606528g^2g_t^4 + 2654208g_3^2g_t^4 - 9870336g_t^6 - 451584g'^4\pi^2 - 12552192g^4\pi^2 + 9216000g'^2g_t^2\pi^2 + 25878528g^2g_t^2\pi^2 + 7077888g_3^2g_t^2\pi^2 + 9289728g_t^4\pi^2 - 15552g'^2g^2(215g_t^2 + 576\pi^2))$

$C_4 = \frac{1}{420864}\Big(-373968g'^8 - 3935808g'^6g^2 - 4327992g'^4g^4 + 1648080g'^2g^6 + 9961704g^8 + 836352g'^6g_3^2 + 380160g'^4g^2g_3^2 + 311040g'^2g^4g_3^2 + 2052864g^6g_3^2 + 3066768g'^6g_t^2 + 4860432g'^4g^2g_t^2 + 5916240g'^2g^4g_t^2 - 11108016g^6g_t^2 - 622080g'^4g_3^2g_t^2 - 1866240g^4g_3^2g_t^2 + 6269280g'^4g_t^4 - 1272672g^4g_t^4 - 10298880g'^4g^2\pi^2 + 12123648g'^2g^4\pi^2 + 4769280g^6\pi^2 - 5971968g'^4g_t^2\pi^2 - 17915904g^4g_t^2\pi^2 - 27648000g'^2g_t^4\pi^2 - 77635584g^2g_t^4\pi^2 - 21233664g_3^2g_t^4\pi^2 + 107495424g_t^6\pi^2 - 15552g'^2g^2g_t^2(80g_3^2 + 351g_t^2 + 768\pi^2)\Big)$

$C_5 = \frac{1}{420864}(-5217399g'^{10} - 7881141g'^8g^2 - 8492370g'^6g^4 - 10152036g'^4g^6 - 973827g'^2g^8 + 19030869g^{10} - 7204032g'^8g_3^2 - 7083648g'^6g^2g_3^2 - 4728960g'^4g^4g_3^2 - 2996352g'^2g^6g_3^2 + 14230080g^8g_3^2 + 270348g'^8g_t^2 + 5768352g'^6g^2g_t^2 + 2429424g'^4g^4g_t^2 + 2604744g'^2g^6g_t^2 - 10589940g^8g_t^2 + 859968g'^6g_3^2g_t^2 + 7967808g'^4g^2g_3^2g_t^2 + 6253632g'^2g^4g_3^2g_t^2 - 18595008g^6g_3^2g_t^2 + 10637568g'^4g_3^4g_t^2 + 5038848g^4g_3^4g_t^2 - 3377956g'^6g_t^4 -$

$1906740g'^4g^2g_t^4 - 669060g'^2g^4g_t^4 + 7341516g^6g_t^4 - 11793408g'^4g_3^2g_t^4 + 10824192g^4g_3^2g_t^4 + 7925760g'^2g_3^4g_t^4 - 14929920g^2g_3^4g_t^4 + 160579584g_3^6g_t^4 + 3317040g'^4g_t^6 + 11294640g^4g_t^6 - 10630656g'^2g_3^2g_t^6 + 5598720g^2g_3^2g_t^6 - 261273600g_3^4g_t^6 + 9328608g'^2g_t^8 + 13366944g^2g_t^8 + 47278080g_3^2g_t^8 - 4012416g_t^{10} - 54768096g'^8\pi^2 - 57553920g'^6g^2\pi^2 - 5095296g'^4g^4\pi^2 + 22635648g'^2g^6\pi^2 - 26420256g^8\pi^2 - 18745344g'^6g_t^2\pi^2 + 21067776g'^4g^2g_t^2\pi^2 - 8875008g'^2g^4g_t^2\pi^2 - 42550272g^6g_t^2\pi^2 + 10616832g'^4g_3^2g_t^2\pi^2 + 6912000g'^4g_t^4\pi^2 - 25132032g^4g_t^4\pi^2 + 76087296g'^2g_3^2g_t^4\pi^2 + 143327232g^2g_3^2g_t^4\pi^2 + 244187136g_3^4g_t^4\pi^2 - 49268736g'^2g_t^6\pi^2 - 40310784g^2g_t^6\pi^2 - 429981696g_3^2g_t^6\pi^2 + 86593536g_t^8\pi^2 - 7776g'^2g^2g_t^2(3024g_3^4 - 369g_t^4 + 64(64g_3^2 - 27g_t^2)\pi^2))$

$$f_1(\xi, M) = 6\ln\frac{M^2}{M_H^2} + \frac{3}{2}\ln\xi - \frac{1}{2}Z\left(\frac{1}{\xi}\right) - Z\left(\frac{c^2}{\xi}\right) - \ln c^2 + \frac{9}{2}\left(\frac{25}{9} - \frac{\pi}{\sqrt{3}}\right)$$

$$f_0(\xi, M) = -6\ln\frac{M^2}{M_Z^2}\left[1 + 2c^2 - 2\frac{M_t^2}{M_Z^2}\right] + \frac{3c^2\xi}{\xi - c^2}\ln\frac{\xi}{c^2} + 2Z\left(\frac{1}{\xi}\right) + 4c^2Z\left(\frac{c^2}{\xi}\right) + \frac{3c^2\ln c^2}{s^2} + 12c^2\ln c^2 - \frac{15}{2}(1 + 2c^2) - 3\frac{M_t^2}{M_Z^2}\left[2Z\left(\frac{M_t^2}{M_Z^2\xi}\right) + 4\ln\frac{M_t^2}{M_Z^2} - 5\right]$$

$$f_{-1}(\xi, M) = 6\ln\frac{M^2}{M_Z^2}\left[1 + 2c^4 - 4\frac{M_t^4}{M_Z^4}\right] - 6Z\left(\frac{1}{\xi}\right) - 12c^4Z\left(\frac{c^2}{\xi}\right) - 12c^4\ln c^2 + 8(1 + 2c^4) + 24\frac{M_t^4}{M_Z^4}\left[\ln\frac{M_t^2}{M_Z^2} - 2 + Z\left(\frac{M_t^2}{M_Z^2\xi}\right)\right]$$

$c^2=\cos^2\theta_W$, $s^2=\sin^2\theta_W$ and

$$Z(z) = \begin{cases} 2A\tan^{-1}\left(\frac{1}{A}\right) & (z > \frac{1}{4}) \\ A\ln\frac{1+A}{1-A} & (z < \frac{1}{4}) \end{cases},$$

$$A = \sqrt{|1 - 4z|}\,.$$